\documentclass[%
 reprint,
superscriptaddress,
 amsmath,amssymb,
 aps,
]{revtex4-2}

\usepackage{graphicx}
\usepackage{dcolumn}
\usepackage{bm}

\usepackage{color, hyperref}
\definecolor{mygrey}{gray}{0.80}
\definecolor{darkblue}{RGB}{8,81,156}
\hypersetup{colorlinks,breaklinks,
            linkcolor=darkblue,urlcolor=darkblue,
            anchorcolor=darkblue,citecolor=darkblue}

\definecolor{super-dark-green}{RGB}{0,69,41}
\definecolor{super-dark-purple}{RGB}{63,0,125}
\definecolor{super-dark-blue}{RGB}{8,48,107}
\definecolor{super-dark-red}{RGB}{165,0,38}

\setcitestyle{super}

\begin{document}

\title{\LARGE Path Dependence in Alchemical Calculations of Water Chemical Potential in Aqueous Electrolytes}

\author{Arlind Kacirani}
\affiliation{%
 Department of Chemical and Environmental Engineering, Yale University, New Haven, CT 06520, USA
}%
\affiliation{%
 Integrated Graduate Program in Physical and Engineering Biology, Yale University, New Haven, CT 06520, USA
}%
\author{Bet\"ul Uralcan}%
\affiliation{%
 Department of Chemical and Environmental Engineering, Yale University, New Haven, CT 06520, USA
}%
\affiliation{%
 Materials Science and Nanoengineering Program, Faculty of Engineering and Natural Sciences, Sabanc{\i} University, Istanbul, T\"urkiye
}%
\author{Amir Haji-Akbari}
\affiliation{%
 Department of Chemical and Environmental Engineering, Yale University, New Haven, CT 06520, USA
}%
\affiliation{%
 Quantitative Biology Institute, Yale University, New Haven, CT 06520, USA
}%
\affiliation{
Wu Tsai Institute, Yale University, New Haven, CT, USA
}

\date{\today}

\begin{abstract}
\noindent 
Accurate calculation of free energies and their derivatives is central to assessing the thermodynamic stability of molecular and particulate systems across length scales. Yet such quantities can be difficult to compute reliably in strongly interacting systems, such as solutions of ionic species in polar solvents. One important example is the chemical potential of water in aqueous electrolytes, which can be estimated through staged particle insertion by gradually coupling an inserted molecule to its environment. Although the resulting insertion free energy should be independent of the alchemical pathway, the order and manner in which van der Waals and electrostatic interactions are activated can strongly affect convergence and, in some cases, yield inconsistent estimates. Here, we examine this issue by calculating water's chemical potential in aqueous KCl solutions using eight alchemical insertion pathways that differ in the extent and order of van der Waals and Coulombic coupling. We find that concurrently activating these interactions, particularly in fully coupled and partially end-coupled protocols, can produce chemically implausible insertion free energies. These anomalies arise from intermediate states in which the inserted water molecule develops strong electrostatic interactions with a chloride ion before sufficient short-range repulsion has been established. In contrast, pathways that activate short-range van der Waals interactions before electrostatics yield more consistent and chemically plausible estimates. These findings  demonstrate that practical alchemical calculations in polar and ionic environments can be highly sensitive to pathway design, underscoring the importance of decoupling short-range and electrostatic interactions in staged insertion alchemical protocols.
\end{abstract}

\maketitle


    \section{Introduction}
    \label{section:intro}

\noindent 
Free energy calculations are central to molecular simulations because they provide a quantitative link between microscopic interactions and experimentally measurable thermodynamic observables. Despite substantial methodological advances\cite{KirkwoodJCP1935, Bennett1976EfficientData,  RomanoMolPhy1979, FrenkelJCP1984, FrenkelPRL1986, Berg1992MulticanonicalTransitions, OrkoulasJCP1994, WangPRL2001, HajiAkbariJCP2011, EspinosaJCP2014, ThaparJCP2015, SulantayCGD2026} and a large body of established practice,\cite{PohorilleJPCB2010, Shirts2013AnCalculations, Hansen2014PracticalReview,  Klimovich2015GuidelinesCalculations, Klimovich2015AGROMACS} accurately determining free energies remains challenging, particularly in complex condensed-phase systems.\cite{HummerJPC1996, BoreschJPCB2003, PalmerNature2014, PalmerJCP2018, SaraviJPCB2021, Kacirani_2024} The primary difficulty is sampling: although free energy is a state function, its evaluation requires integration over high-dimensional, correlated configurational spaces that often exhibit slow relaxation. Entropic contributions, in particular, cannot be recovered from simple trajectory averages, but instead require extensive exploration of the relevant ensemble. Inadequate sampling therefore propagates directly into systematic biases and numerical instabilities, often leading to poor convergence or even simulation failure.\cite{Beveridge1989FreeSystems,Pitera2002ACalculations}

Over the past several decades, a wide variety of techniques have been developed to estimate and analyze free energies from atomistic simulations. Prominent examples include exponential averaging (\textsc{Exp})-- also known as the \textit{Zwanzig relationship},\cite{Zwanzig1954HighTemperatureGases}  thermodynamic integration\cite{Resat1993StudiesPath, Jorge2010EffectIntegration} (\textsc{Ti}), the weighted histogram analysis method\cite{Ferrenberg1989OptimizedAnalysis} (\textsc{Wham}), Bennett’s acceptance ratio\cite{Bennett1976EfficientData} (\textsc{Bar}), and its multistate extension, \textsc{Mbar}.\cite{Shirts2008StatisticallyStates} Many of these approaches, including \textsc{Ti} and \textsc{Wham}, rely on constructing a continuous thermodynamic pathway connecting a reference state to a target state. Their reliability, however, strongly depends on sufficient configurational overlap between configurational ensembles of neighboring states.

One important application of these methods is the calculation of the chemical potential of a molecular species in solution. The solvent chemical potential, particularly its excess component, plays a central role in determining solution thermodynamics and phase behavior.\cite{Koop2000WaterSolutions, LuoJPCL2010,  SauterJCTC2016, EspinosaJPCL2017} One of the earliest and conceptually simplest strategies for estimating chemical potentials is the \emph{Widom particle insertion} method,\cite{Wldom1963SomeFluids} in which the chemical potential is obtained from $-\beta^{-1}\ln\langle \exp[-\beta \Delta U]\rangle$, where $\Delta U$ is the change in potential energy upon insertion of a test particle into the system. This approach performs reasonably well in moderately dense fluids.\cite{Frenkel1996UnderstandingApplications} In dense liquids, however, especially those with strong directional interactions such as water and aqueous electrolyte solutions, random test particle insertions often result in large increases in potential energy arising from strong short-range repulsions (e.g.,~due to the disruption of hydrogen-bond networks). Consequently, the exponential average becomes dominated by rare favorable insertions, leading to slow convergence and large statistical uncertainties.\cite{MooreJCP2011, PeregoEurPhysJSpecTop2016, QinJPCB2016}

To overcome these limitations, several alternative approaches have been developed, including biased insertion, particle-deletion schemes, overlap-based estimators, expanded ensembles, and related enhanced-sampling strategies.\cite{ShingMolPhys1982, BoulougourisMolPhys1999, DelgadoJCP2005, NeimarkJCP2005, QinJCTC2013, PeregoEurPhysJSpecTop2016, QinJPCB2016} Among the most widely used are staged-insertion\cite{MonPRA1985} and alchemical-transformation\cite{SquireJCP1969} methods. In these approaches, one or more coupling parameters are used to gradually activate the interactions between the inserted molecule and its environment, thereby transforming a noninteracting reference state into the fully interacting target state. Simulations at intermediate coupling parameters provide ensemble averages of the Hamiltonian gradient with respect to the coupling variables, which are then integrated along the alchemical pathway to obtain the free energy difference.

Because free energy is a state function, the computed free energy difference should, in principle, be independent of the chosen alchemical path, provided that all intermediate states are sampled reversibly and with adequate phase-space overlap. In practice, however, the choice of the pathway can strongly affect both convergence and numerical accuracy. Several rigorous frameworks have therefore been developed to reduce path-dependent errors and improve estimator efficiency, including solvation-theory-based approaches for identifying low-variance transformations,\cite{Pham2011IdentifyingPhase} as well as more formal treatments based on thermodynamic geometry\cite{Shenfeld2009MinimizingSimulations} and optimal transport theory.\cite{Decherchi2023OptimalEstimation} Although appealing, these methods typically require detailed characterization of the coupling-parameter-dependent configurational ensemble, which can make them difficult to deploy broadly.

In this work, we examine staged-insertion calculations of the solvent excess chemical potential in aqueous electrolyte solutions. We show that the computed chemical potentials can depend strongly on the chosen coupling path, even though the underlying free energy is formally path independent. In particular, certain coupling schemes generate poorly sampled, unphysical states that introduce systematic biases and lead to chemically implausible estimates. By analyzing these discrepancies at the molecular level, we identify their microscopic origin and clarify how the ordering of short-range and electrostatic interactions controls the reliability of solvent chemical-potential calculations in electrolyte solutions. 

The remainder of this paper is organized as follows. Sections~\ref{section:method:MD} and~\ref{section:method:insertion} describe the molecular dynamics simulations and free energy calculation methodology, respectively. Section~\ref{section:results} presents the simulation results, and Section~\ref{section:conclusions} summarizes the main conclusions while providing some forward-looking observations.

\section{Methods}

\begin{figure}
    \centering
    \includegraphics[width=0.5\textwidth]{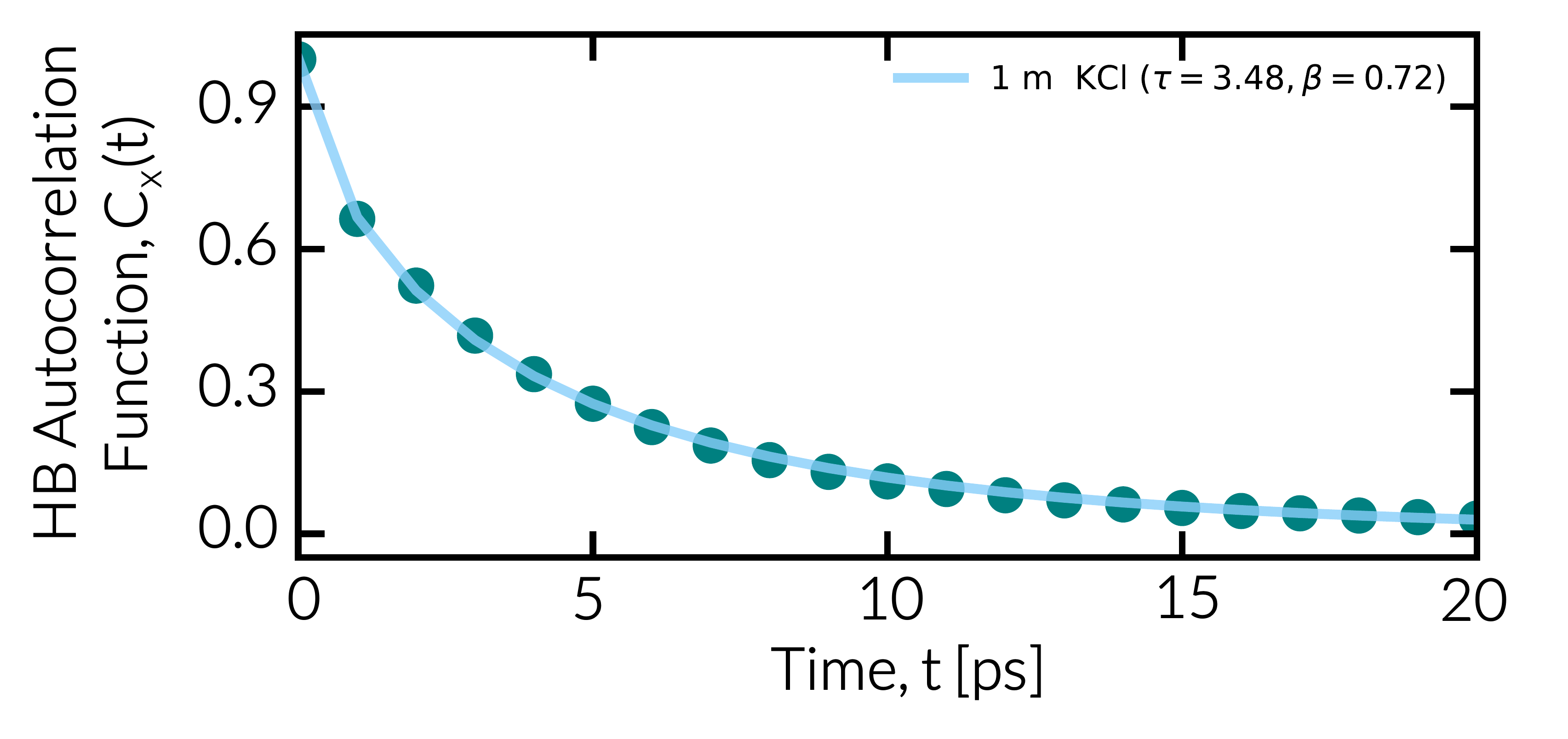}
    \caption{\label{fig.2} Hydrogen bond autocorrelation function, $C_{HB}(t)$ (dark green circles) computed for a 1~m aqueous KCl solution with the stretched exponential fit depicted in cyan.}
\end{figure}

\begin{figure*}
    \centering
    \includegraphics[width=\textwidth]{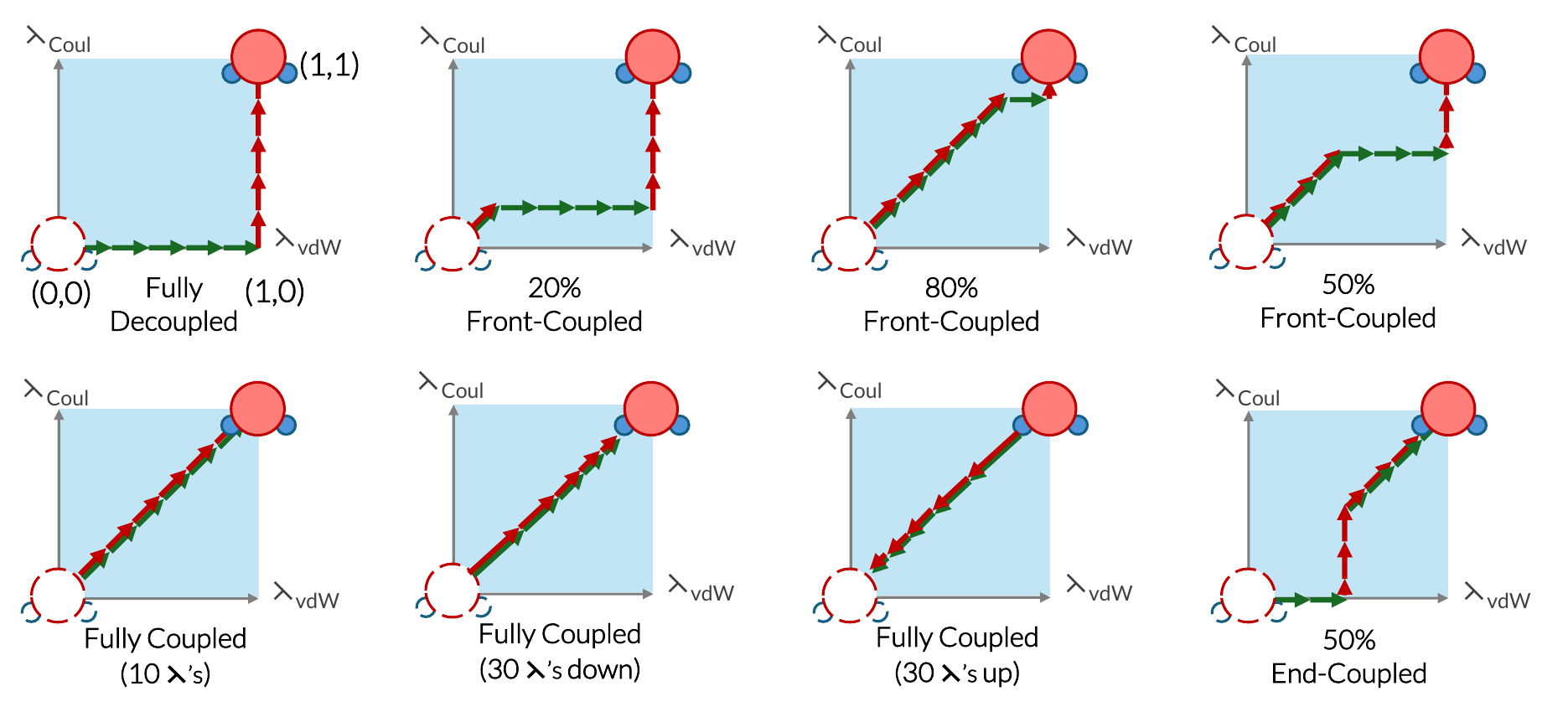}
    \caption{\label{fig.1}Schematic representation of paths with different coupling protocols within the $(\lambda_{\text{vdW}},\lambda_{\text{Coul}})$ space. Arrows denote the direction of alchemical paths.}
\end{figure*}

\subsection{Molecular Dynamics Simulations}
\label{section:method:MD}

\noindent
In this work, we consider an aqueous $1\,\text{m}$ solution of KCl, with water molecules  represented using the \textsc{Tip4p-Ew} model\cite{Horn2004DevelopmentTIP4P-Ew} while potassium and chloride ions are described by the Joung--Cheatham (\textsc{Jc}) force-field.\cite{SukJoung2008DeterminationSimulations} 
All \textsc{Md} simulations are performed using the open-source \textsc{Gromacs} package, Version 2021.5,\cite{ABRAHAM201519} with Newton's equations of motion integrated using the leapfrog algorithm\cite{HockneyMethodsComputPhys1970} with a time step of 2~fs. Two types of \textsc{Md} simulations are conducted, both in the isothermal isobaric (\textsc{Npt}) ensemble. To generate starting configurations for staged particle insertion, we use conventional \textsc{Md} simulations wherein temperature and pressure are controlled using the Nos\'{e}-Hoover thermostat \cite{NoseMolPhys1984, HooverPhysRevA1985} and the Parrinello-Rahman barostat,\cite{Parrinello1981PolymorphicMethod} with time constants of 0.4~ps and 2~ps, respectively. In staged particle insertion simulations discussed in detail in Section \ref{section:method:insertion}, stochastic dynamics is employed with a Langevin thermostat and a C-rescale barostat\cite{BernettiJCP2020} used for controlling temperature and pressure, respectively. Simulation boxes are comprised of 6,899 water molecules and 129 ion pairs, and are periodic in all three directions. Long-range electrostatic interactions are treated using the smooth particle mesh Ewald\cite{Essmann1995AMethod}  (\textsc{Pme}) method with a grid spacing of 0.1~nm, while all short-range interactions are truncated at 1~nm. The rigidity constraints within the \textsc{Tip4p-Ew} model are enforced using the linear constraint solver (\textsc{Lincs}).\cite{Hess_2007}

To generate equilibrated configurations for stochastic dynamics, conventional \textsc{Npt} \textsc{Md} trajectories are initiated from ten randomly generated configurations prepared using the \texttt{gmx genion} command, which randomly replaces solvent molecules with K$^+$ or Cl$^-$ ions. To determine the optimal equilibration and production times, we compute the hydrogen bond (HB) autocorrelation function\cite{LuzarNature1996} defined as,
\begin{eqnarray}
C_{HB}(t) = \frac
{\left\langle \sum_{i<j} h_{ij}(t)h_{ij}(0)\right\rangle}
{\left\langle \sum_{i<j} h_{ij}(0)\right\rangle}.
\end{eqnarray}
Here, $h_{ij}(t)$ is an indicator function that is 1 if molecules $i$ and $j$ are hydrogen bonded at time $t$ and zero otherwise. To determine whether two water molecules are hydrogen bonded, we use the geometric criterion outlined in Van der Spoel \textit{et al}.\cite{van_der_Spoel_2006} The  $C_{HB}(t)$ computed for a $1\,\text{m}$ KCl solution is shown in Fig.~\ref{fig.2}, and is fitted to a stretched exponential function of the form,
\begin{equation} \label{eq.1}
C_{HB}(t) = A e^{(-t/\tau)^\beta},
\end{equation}
where $A$ is the exponential prefactor, $\tau$ is the hydrogen bond relaxation time, and $\beta$ is the dimensionless stretch exponent. Since the hydrogen-bond relaxation time is on the order of picoseconds (i.e.,~3.48 ps), we argue that a minimum of 100 relaxation times is needed for proper sampling, amounting to approximately 350~ps. In practice, each state is sampled for a minimum of $660$~ps, with the first 100~ps discarded prior to estimating the ensemble averages.

\subsection{Free Energy Calculations} \label{methods: free-energy-calc}
\label{section:method:insertion}

\noindent
As stated above, staged particle insertion is a generalization of the Widom particle insertion method,\cite{Wldom1963SomeFluids} which expresses the chemical potential of a given species (in our case water) as,
\begin{eqnarray}
\mu_w &=& \left(
\frac{\partial G}{\partial n_w}
\right)_{P,T,n_{j\neq w}} \approx G(n_w+1)-G(n_w) \notag\\
&=& -k_BT \ln
\frac
{\Xi(n_w+1,n_{j\neq w},P,T)}
{\Xi(n_w,n_{j\neq w},P,T)}.
\end{eqnarray}
Here, $\Xi(\cdot)$ is the isothermal-isobaric partition function, and $k_B$ is the Boltzmann constant. In staged particle insertion, the process of adding a new water molecule to the system is carried out gradually using a system Hamiltonian, $\mathcal{H}(\Gamma,\pmb\lambda)$ that depends on a collection of coupling parameters $\pmb\lambda \equiv (\lambda_1,\cdots,\lambda_m)$. (Here, $\Gamma$ denotes the configurational degrees of freedom of the system.) If $\mathcal{H}(\Gamma,\pmb\lambda)$ does not possess an explicit dependence on  density, it is easy to demonstrate that:
\begin{eqnarray}\label{eq:DeltaG-int}
\nabla_{\pmb\lambda}G &=& -k_BT\nabla_{\pmb\lambda}\ln \int 
e^{-\beta\left[\mathcal{H}(\Gamma,\pmb\lambda)+PV\right]}\,d\Gamma\,dV \notag\\
&=& \frac
{\displaystyle\int \nabla_{\pmb\lambda} \mathcal{H}\,e^{-\beta\left[\mathcal{H}(\Gamma,\pmb\lambda)+PV\right]}\,d\Gamma\,dV}
{\displaystyle\int e^{-\beta\left[\mathcal{H}(\Gamma,\pmb\lambda)+PV\right]}\,d\Gamma\,dV} = \left\langle
\nabla_{\pmb\lambda}\mathcal{H}
\right\rangle_{\pmb\lambda}
\label{eq:grad-G-lambda}
\end{eqnarray}
The free energy difference between two states is then estimated by choosing a continuous path $\mathcal{C}$, within the $\pmb\lambda$ space and numerically integrating \eqref{eq:grad-G-lambda}:
\begin{eqnarray}\label{eq:Delta-G-integrated}
G(\pmb\lambda_2) - G(\pmb\lambda_1) &=& \int_{\mathcal{C}} d\pmb\lambda\cdot  \left\langle
\nabla_{\pmb\lambda}\mathcal{H}
\right\rangle_{\pmb\lambda}
\end{eqnarray}
provided that the path is sampled reversibly.

In the case of adding a water molecule, its interactions with the rest of the system are continuously tuned through two coupling parameters, $\lambda_{\text{vdW}}$ and $\lambda_{\text{Coul}}$, which determine the extent and softness of van der Waals and Coulombic interactions, respectively. More precisely, the system Hamiltonian is expressed as,
$$
\mathcal{H} = \mathcal{H}(n_w,n_+,n_-) + \lambda_{\text{vdW}} U_{\text{vdW}}( \lambda_{\text{vdW}}) + \lambda_{\text{Coul}} U_{\text{Coul}}( \lambda_{\text{Coul}})
$$
Here, $n_+=n_-$  correspond to the number of K$^+$ and Cl$^-$ ions in the system. $U_{\text{vdW}}(\lambda_{\text{vdW}})$ and $U_{\text{Coul}}( \lambda_{\text{Coul}})$ comprise sums of Lennard-Jones (\textsc{Lj}) and Coulombic pair interactions between the inserted molecule and the rest of the system, both scaled according to the soft-core versions of Beutler~\emph{et al.}\cite{Beutler1994AvoidingSimulations}  to avoid singularities during the insertion process:
{\small
\begin{eqnarray}
\phi_{ij}^{\textsc{Lj}} (r;\lambda_{\text{vdW}}) &=& 4\epsilon_{ij}\Bigg\{
\frac{1}{\left[\alpha_{\textsc{Lj}}(1-\lambda_{\text{vdW}})^2+\left(r/\sigma_{ij}\right)^6\right]^2}\notag\\
&& - \frac{1}{\alpha_{\textsc{Lj}}(1-\lambda_{\text{vdW}})^2+\left(r/\sigma_{ij}\right)^6} \Bigg\}\label{eq:softcore-LJ}\\
\phi_{ij}^{\textsc{Coul}} (r;\lambda_{\text{Coul}}) &=& 
\frac{q_iq_j}{4\pi\epsilon_0\sigma_{ij}\left[\alpha_{\textsc{Coul}}(1-\lambda_{\text{Coul}})^2+\left(r/\sigma_{ij}\right)^6\right]^{1/6}}\notag\\
&&\label{eq:softcore-Coul}
\end{eqnarray}
}
Here, $\epsilon_{ij}$ and $\sigma_{ij}$ correspond to the \textsc{Lj} pair interaction parameters, $\epsilon_0$ is the vacuum permittivity, and $\alpha$ is the soft-core parameter (in this work $\alpha_{\textsc{Lj}} = \alpha_{\textsc{Coul}}= 1$).

Staged particle insertion is conducted by choosing an appropriate path between $\pmb\lambda_1=(0,0)=\pmb0$-- i.e.,~where the inserted molecule does not interact with the rest of the system-- and $\pmb\lambda_2=(1,1)=\pmb1$ where the interactions are fully active. We will refer to the $\Delta{G}$ obtained from this process as the \emph{insertion free energy}, $\Delta G_{\text{ins}}$. Alternatively, this free energy difference is sometimes referred to as \emph{solvation free energy} as the inserted molecule is solvated by the existing entities in the system, although the latter terminology can be potentially confusing for inserting a solvent molecule.

To examine how the choice of the coupling pathway impacts the computed free energy difference, we consider several coupling schemes that differ in the degree of simultaneous scaling between electrostatic and van der Waals interactions (see Fig.~\ref{fig.1}). In addition to a \emph{fully decoupled} path in which \textsc{Lj} interactions are fully turned on prior to activating electrostatic interactions, namely $(0,0)\rightarrow(1,0)\rightarrow(1,1)$, we consider the following seven partially or fully coupled paths:
\begin{itemize}
\item three \emph{fully coupled} paths (along the $(0,0)\rightarrow(1,1)$ diagonal): a linear path comprised of ten evenly distributed $\pmb\lambda$'s, and two non-linear (concave-up and down) paths in which 30 $\pmb\lambda$'s are spaced quadratically along the diagonal;
\item three partially \emph{front-coupled} paths comprised of 20 evenly distributed $\pmb\lambda$'s, where 20\%, 50\%, and 80\% of the path involves simultaneously activating \textsc{Lj} and electrostatic interactions, prior to switching to a decoupled path;
\item an end-coupled path that follows the reverse order of the 50\% front-coupled path, namely $(0,0)\rightarrow(0.5,0)\rightarrow(0.5,0.5)\rightarrow(1,1)$.
\end{itemize}
Taken together, these paths probe how the relative ordering and extent of coupling between \textsc{Lj} and electrostatic interactions influence the sampling efficiency and convergence of staged particle insertion. 

\begin{figure*}
    \centering
    \includegraphics[width=\textwidth]{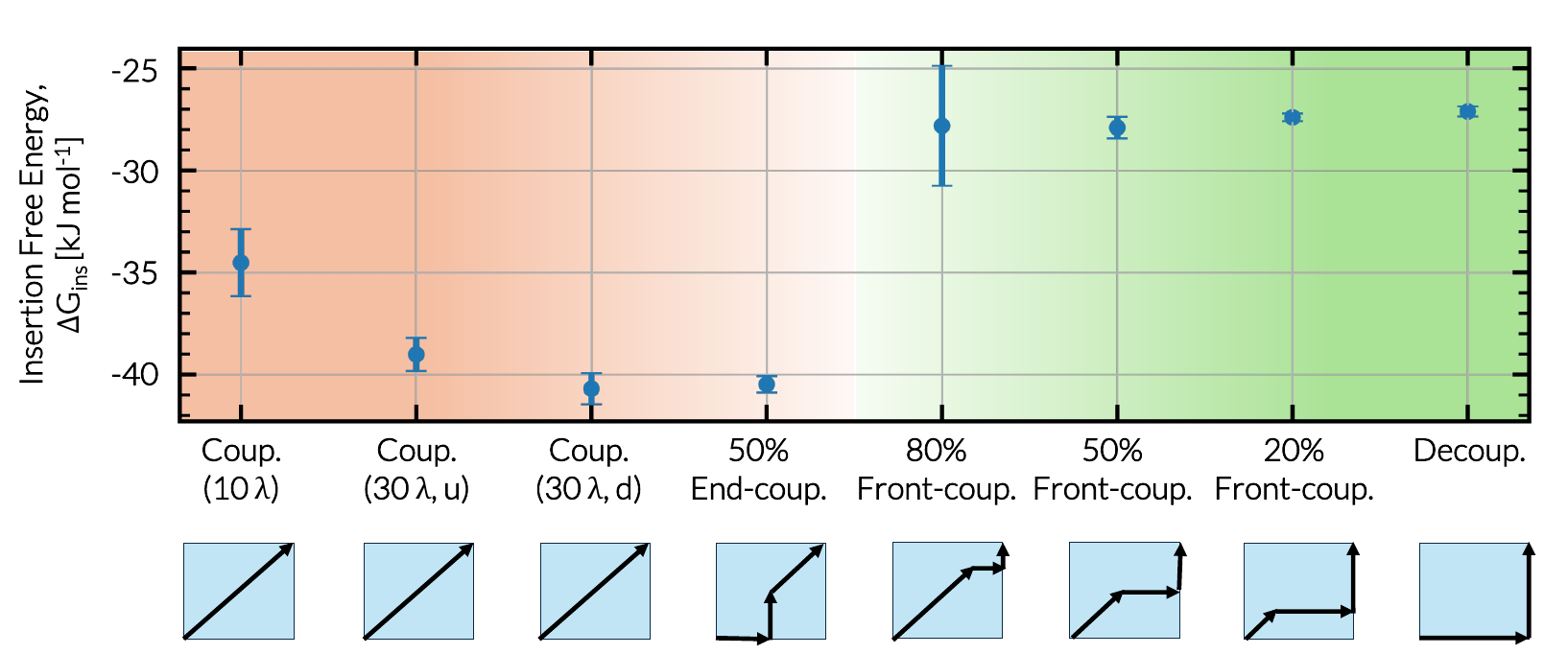}
    \caption{\label{fig.3} Water excess chemical potential,~i.e.,~excess insertion free energy, obtained from paths with different coupling protocols.}
\end{figure*}

In principle, the  insertion free energy can be computed by numerically integrating Eq.~\eqref{eq:DeltaG-int}. To achieve enhanced numerical accuracy considering the discretized nature of the paths, however, we estimate $\Delta{G}$ as the sum of $\Delta{G}$'s between successive states. We determine the latter using the \emph{Bennett acceptance ratio}\cite{Bennett1976EfficientData} (\textsc{Bar}) method as implemented in \textsc{Gromacs} through the \texttt{gmx bar} utility. The \textsc{Bar} method estimates the free energy difference between the states $A$ and $B$ by sampling both states and iteratively solving the following algebraic equation:
\begin{equation} \label{eq.5}
\begin{split}  
1 = \frac{\langle f[\beta (-\mathcal{H}_B + \mathcal{H}_A - \Delta G)] \rangle_B}{\langle f[\beta(\mathcal{H}_B - \mathcal{H}_A + \Delta G)] \rangle_A}
\end{split}
\end{equation}
for $\Delta{G}$. Here, $f(x) = 1/(1 + e^x)$  and we assume that an equal number of independent samples is generated at each state. In the context of particle insertion, $A$ and $B$ correspond to successive points along the paths outlined above. Each reported insertion	 free energy is obtained from a minimum of ten particle insertion simulations initiated from independent starting configurations. 

   
    \section{Results and Discussion}
    \label{section:results}

    \subsection{Path Dependence of the Computed Insertion Free Energies}

\noindent
Figure~\ref{fig.3} and Table~\ref{table.1} present the insertion free energies computed using the different coupling schemes. Surprisingly, the resulting estimates are not statistically consistent across pathways, despite the fact that the Gibbs free energy is a state function and, in principle, should be independent of the thermodynamic path. A more detailed analysis shows that both the degree of coupling between van der Waals and electrostatic interactions and the sequence in which these interactions are introduced substantially affect the final estimate.

In particular, coupling schemes in which \textsc{Lj} interactions are fully activated before the complete introduction of electrostatic interactions---namely, the fully decoupled path and three of the four partially coupled paths---produce insertion free energies that are chemically plausible and statistically well converged. By contrast, protocols in which \textsc{Lj} and electrostatic interactions are activated concurrently near the end of the transformation---specifically, the fully coupled scheme with 10 linearly spaced $\lambda$ windows, the 30-$\lambda$ concave-up and concave-down paths, and the 50\% end-coupled path---yield insertion free energies that are chemically implausible, severely underestimated relative to the experimental value of $-26.6$~kJ mol$^{-1}$,\cite{Han_2015} and characterized by substantially larger statistical uncertainties. These estimates are not merely noisy; they are systematically biased and statistically distinct from those obtained using fully decoupled or partially front-coupled protocols.

A particularly striking example is provided by the comparison between the 50\% front-coupled and 50\% end-coupled paths. Although these schemes differ only in the order in which \textsc{Lj} and Coulombic interactions are activated, the resulting free energy estimates differ by more than 12 kJ mol$^{-1}$ (Table~\ref{table.1}). This pronounced discrepancy indicates that not only the extent of coupling but also its order can significantly affect the computed insertion free energy, highlighting apparent path dependence arising from incomplete sampling or methodological artifacts.

\begin{table} 
\centering
\caption{\label{table.1}Insertion free energy estimates obtained from paths with different coupling schemes. Reported uncertainties correspond to standard errors.}
\begin{tabular}{ l | c } 
 \hline\hline
 \bf{Path} & \bf{$\Delta G$ [kJ mol$^{-1}$]}  \\ 
 \hline
  coupled (linear) & $-34.518\pm1.645$ \\ 
  coupled (concave-up) & $-39.017\pm0.815$ \\ 
  coupled (concave-down) & $-40.696\pm0.763$ \\  
  fully decoupled  & $-27.110\pm0.252$ \\ 
  80\% front-coupled & $-27.813\pm2.938$ \\ 
  50\% end-coupled & $-40.480\pm0.399$ \\ 
  50\% front-coupled & $-27.903\pm0.526$ \\ 
  20\% front-coupled & $-27.396\pm0.188$ \\ \hline 
\end{tabular}
\end{table}

\begin{figure}
    \centering
    \includegraphics[width=0.5\textwidth]{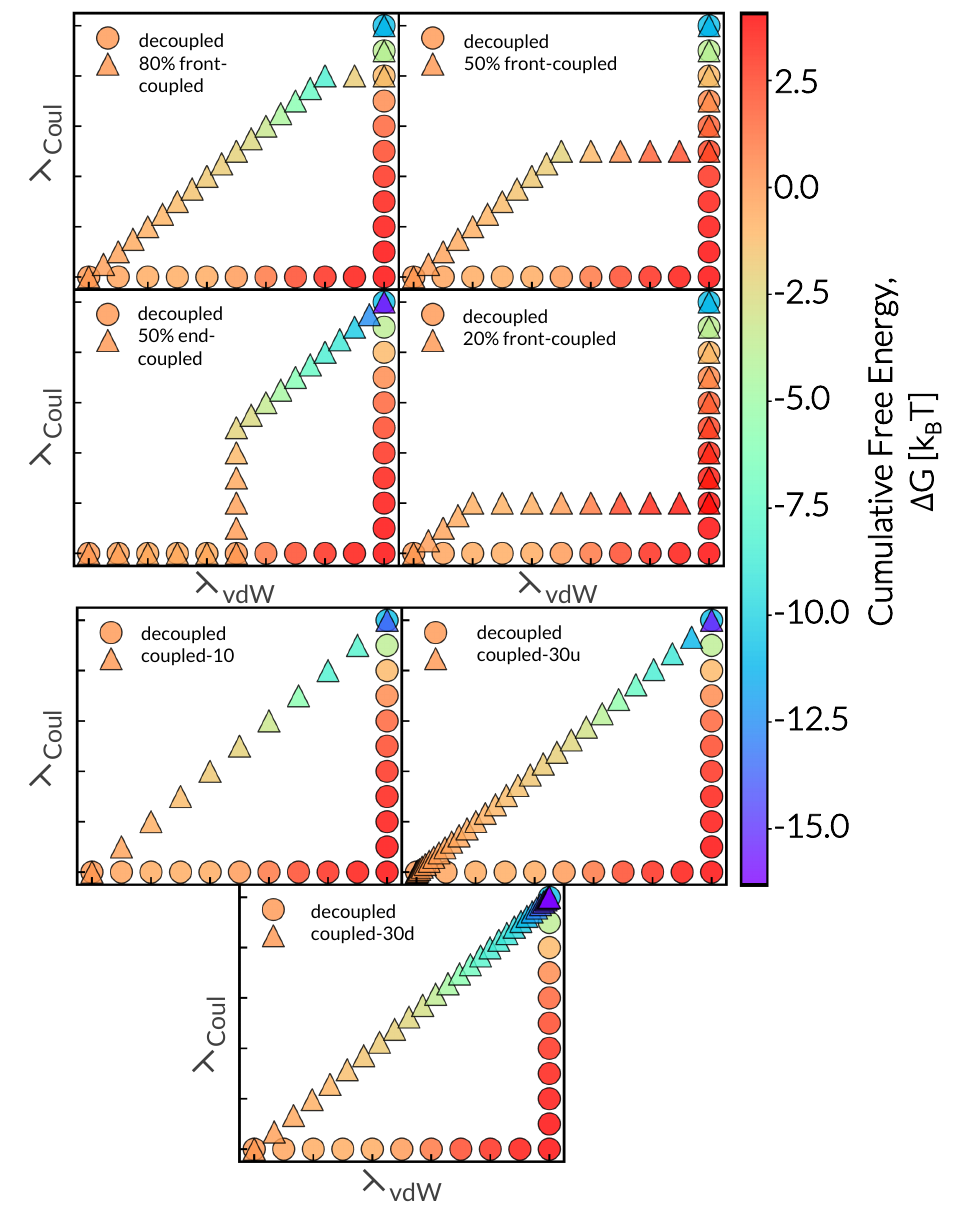}
    \caption{\label{fig.4}Cumulative insertion free energy, $G(\lambda_{\mathrm{vdW}},\lambda_{\mathrm{Coul}})-G(0,0)$, evaluated along the fully decoupled path (circles) and along paths with different levels of coupling (triangles).}
\end{figure}

Motivated by the observed path-dependent underestimation of the free energy, we examine the evolution of the cumulative insertion free energy, $\Delta G_{\mathrm{ins}}$, in the two-dimensional $(\lambda_{\mathrm{vdW}}, \lambda_{\text{Coul}})$ coupling space. Figure~\ref{fig.4} compares the cumulative free energy along the fully decoupled reference path (circles) with those obtained from alternative coupling protocols (triangles), shown for the partially coupled paths in the top four panels and the fully coupled paths in the bottom three panels. In all panels, the color scale represents the relative value of $\Delta G$ in units of $k_BT$ at each intermediate state.

The deviations are localized predominantly in the late stages of the transformation,~i.e.,~as the system approaches the fully interacting molecule. In the three fully coupled protocols, discrepancies in this region lead to a systematic underestimation of the total insertion free energy by approximately $5$–$7\,k_BT$. A comparable deviation is also observed for the 50\% end-coupled path, despite its partial separation of interactions. This behavior further underscores the sensitivity of the final free energy estimate to the direction and sequence of interaction activation, indicating that the cumulative error arises primarily from inadequate sampling or enhanced fluctuations in the strongly interacting regime.

\subsection{Unphysical Water-Ion Contacts due to Weak van der Waals Interactions}

\begin{figure}
    \centering
    \includegraphics[width=0.46\textwidth]{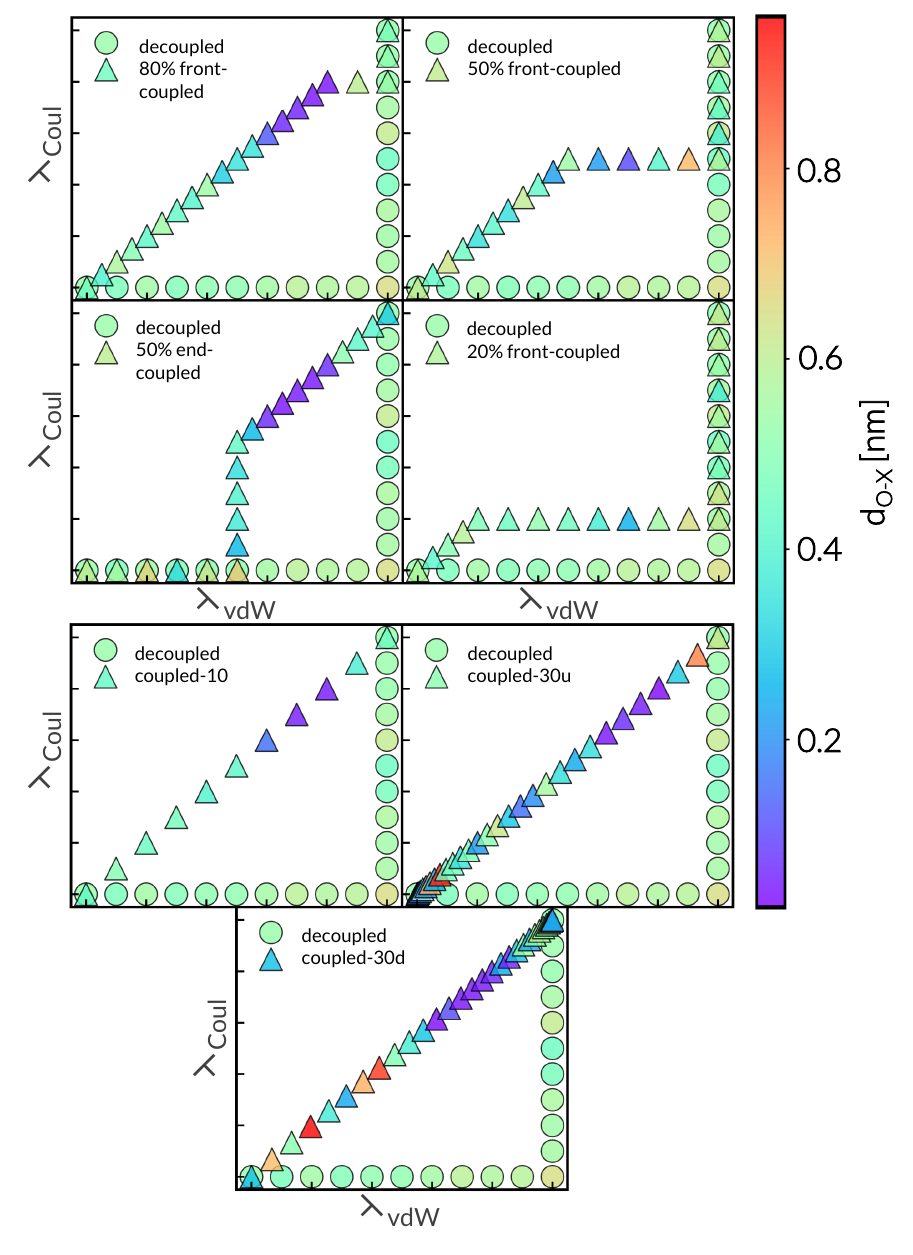}
    \caption{\label{fig.5} Evolution of $d_{\mathrm{O}-X}$, the minimum distance between the inserted water oxygen and the nearest ion in the $(\lambda_{\mathrm{vdW}},\lambda_{\mathrm{Coul}})$ space for the fully decoupled pathway (circles) and pathways with different levels of coupling (triangles).
    }
\end{figure}

\noindent
To identify the microscopic origin of the anomalous free energy estimates, we examine whether the problematic alchemical pathways also generate any anomalous structural features. We focus on
$d_{\mathrm{O}-X}$, defined as the minimum distance between the oxygen atom of the inserted water molecule and any ion in the system. Figure~\ref{fig.5} shows the evolution of $d_{\mathrm{O}-X}$ along pathways with different electrostatic and van der Waals coupling protocols. The pathways that yield chemically implausible or high-variance insertion free energies also exhibit anomalously small values of $d_{\mathrm{O}-X}$. With the exception of the 20\% and 50\% front-coupled pathways, all partially or fully coupled pathways sample configurations with $d_{\mathrm{O}-X}<0.2~\mathrm{nm}$, far below distances expected for physically admissible water--ion contacts in the bulk solution. These short contacts occur primarily near the end of the alchemical path, coinciding with the region in which the corresponding free energy estimates begin to diverge. Visual inspection confirms that, in these configurations, the inserted water molecule can become partially embedded within the van der Waals volume of a nearby monoatomic ion.

\begin{figure*}
    \centering
    \includegraphics[width=.9\textwidth]{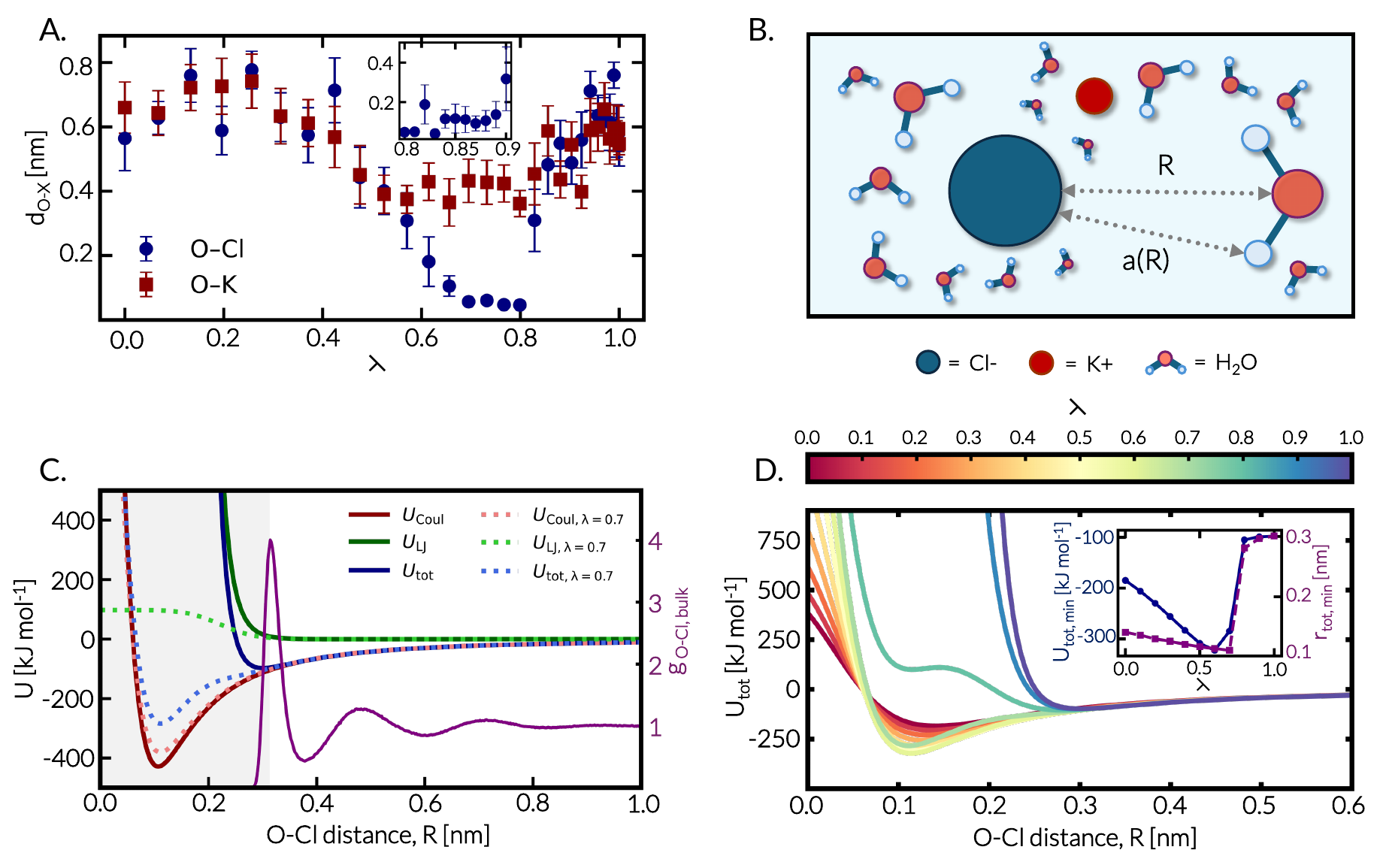}
    \vspace{-10pt}
    \caption{\label{fig.6}
    (A) Minimum oxygen--ion distance, $d_{\mathrm{O}-X}$, as a function of $\lambda_{\mathrm{vdW}}$ for $X\in\{\mathrm{K}^{+},\mathrm{Cl}^{-}\}$. The inset shows $d_{\mathrm{O}-\mathrm{Cl}}$ over a narrower range of $\lambda$ values for which additional simulations were conducted. (B) Schematic geometry of the inserted water molecule relative to the nearest $\mathrm{Cl}^{-}$ ion. 
    (C) Coulombic contribution, $U_{\textsc{Coul}}$ (red), Lennard-Jones contribution, $U_{\textsc{Lj}}$ (green), and total interaction, $U_{\mathrm{tot}}$ (blue), as functions of the chloride--oxygen distance $R$ for $\lambda_{\mathrm{vdW}}=1$ (solid) and $\lambda_{\mathrm{vdW}}=0.7$ (dashed). The oxygen--chloride radial distribution function, $g_{\mathrm{O}-\mathrm{Cl}}(R)$, is shown on the right axis.
    (D) The evolution of $U_{\text{tot}}$~vs.~$R$ for different $\lambda$'s, with the inset depicting the depth and locus of its first minimum vs.~$\lambda$.
    }
\end{figure*}

Further insights are obtained from the fully coupled concave-up pathway. Figure~\ref{fig.6}A shows the minimum oxygen--potassium and oxygen--chloride distances as functions of $\lambda_{\mathrm{vdW}}$. The oxygen--potassium distance remains within a physically reasonable range throughout the transformation. By contrast, $d_{\mathrm{O}-\mathrm{Cl}}$ becomes anomalously small for $0.6 \lesssim \lambda_{\mathrm{vdW}} \lesssim 0.8$, indicating that the dominant structural artifact is associated with water--chloride contacts. This behavior can be rationalized from the local geometry of the inserted water molecule relative to the nearest chloride ion, illustrated schematically in Fig.~\ref{fig.6}B. Using Eqs.~\eqref{eq:softcore-LJ} and \eqref{eq:softcore-Coul}, the Lennard-Jones and Coulombic contributions to the interaction between the inserted water molecule and the nearest chloride ion can be written as
\begin{eqnarray}
U_{\textsc{Lj}}(R,\lambda_{\mathrm{vdW}}) &=&
\phi_{\mathrm{O}-\mathrm{Cl}}^{\textsc{Lj}}(R,\lambda_{\mathrm{vdW}}), \notag\\
U_{\textsc{Coul}}(R,\lambda_{\mathrm{vdW}}) &=&
\phi_{\mathrm{O}-\mathrm{Cl}}^{\textsc{Coul}}(R,\lambda_{\mathrm{vdW}})
-
\phi_{\mathrm{H}-\mathrm{Cl}}^{\textsc{Coul}}\!\left[a(R),\lambda_{\mathrm{vdW}}\right],
\notag
\end{eqnarray}
where $R$ is the chloride--oxygen distance and $a(R)$ is the corresponding chloride--hydrogen distance for the geometry shown in Fig.~\ref{fig.6}B. Figure~\ref{fig.6}C compares $U_{\textsc{Lj}}$, $U_{\textsc{Coul}}$, and their sum,
$U_{\mathrm{tot}}=U_{\textsc{Lj}}+U_{\textsc{Coul}}$, as functions of $R$ for
$\lambda_{\mathrm{vdW}}=1$ and $\lambda_{\mathrm{vdW}}=0.7$. At full van der Waals coupling, the minimum of $U_{\mathrm{tot}}$ coincides with the first peak of the oxygen--chloride radial distribution function, $g_{\mathrm{O}-\mathrm{Cl}}(R)$, as expected for a physically realistic contact. At $\lambda_{\mathrm{vdW}}=0.7$, however, the weakened repulsive core is insufficient to counterbalance the electrostatic attraction, and the minimum of $U_{\mathrm{tot}}$ shifts to substantially smaller $R$ values. This shift explains the formation of unphysically short O--Cl contacts along the coupled pathways.

This analysis also explains why the most severe artifacts do not necessarily occur at  smaller values of $\lambda_{\mathrm{vdW}}$ where the potential is even softer, and therefore less physical. As shown in Fig.~\ref{fig.6}D, intermediate values of $\lambda_{\mathrm{vdW}}$ produce a particularly unfavorable combination: the repulsive core is sufficiently weakened to permit close approach, while the electrostatic attraction remains strong enough to stabilize the contact. The inset of Fig.~\ref{fig.6}D shows that, over this intermediate range, the minimum of $U_{\mathrm{tot}}(R)$ becomes deeper and shifts to smaller $R$ values. These conditions create a local minimum that can trap water--chloride pairs at unrealistically short separations, leading to large fluctuations and biased free energy estimates.

\begin{figure}
    \centering
    \includegraphics[width=.45\textwidth]{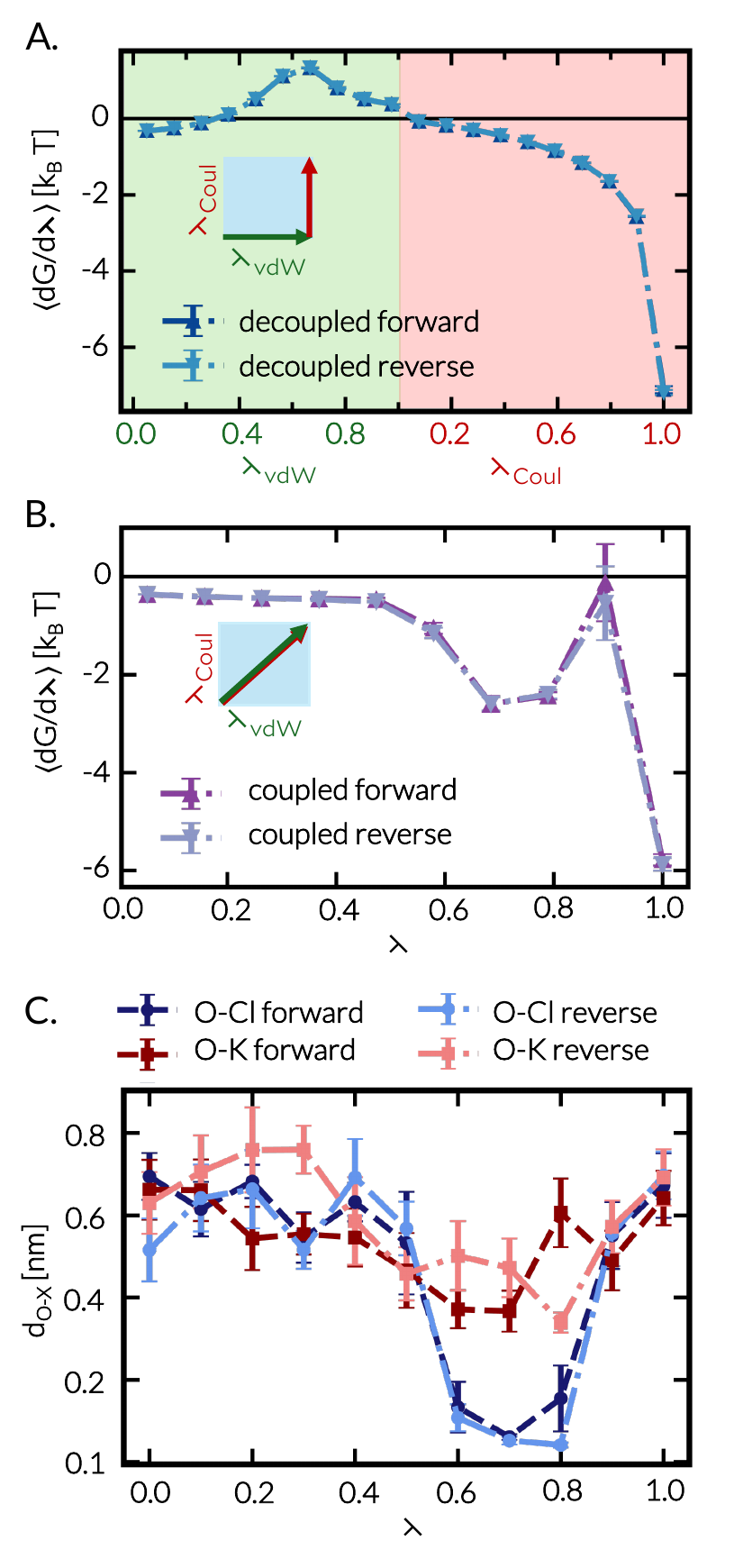}
    \caption{\label{fig.7}
    Comparison of forward and reverse integration along the fully decoupled and the $10-\lambda$ fully coupled pathway. (A-B) $\langle dG/d\lambda \rangle$ vs.~$\lambda$ for (A) the fully decoupled pathway, and (B) the 10-$\lambda$ coupled pathway; and 
    (C) $d_{\mathrm{O}-X}$ as functions of $\lambda$ for forward and reverse integration of  the 10-$\lambda$ coupled pathway.
    }
\end{figure}

We next ask whether these trapped contacts produce detectable hysteresis in the alchemical integration. To this end, we compare forward and reverse integration along the constructed path. As shown in Fig.~\ref{fig.7}A, the decoupled path, which does not produce such artifacts, exhibits no discernible hysteresis between the forward and reverse directions, as reflected in the corresponding $dG/d\lambda$ profiles. Interestingly, the coupled path shows no appreciable forward--reverse hysteresis either (Fig.~\ref{fig.7}B), although the large uncertainties near $\lambda\approx0.8$ point to poor sampling. The same conclusion is supported by the forward and reverse profiles of $d_{\mathrm{O}-\mathrm{Cl}}$ and $d_{\mathrm{O}-\mathrm{K}}$, which are broadly consistent with one another (Fig.~\ref{fig.7}C). Thus, the artifact does not manifest primarily as conventional forward--reverse hysteresis. Rather, it appears to arise from intermittent trapping in deep local minima. Consistent with this interpretation, additional sampling at intermediate $\lambda_{\mathrm{vdW}}$ values produces large fluctuations in $d_{\mathrm{O}-\mathrm{Cl}}$, rather than convergence toward a smooth, reproducible trend, as shown in the inset of Fig.~\ref{fig.6}A.

\begin{figure}
    \centering
    \includegraphics[width=0.3\textwidth]{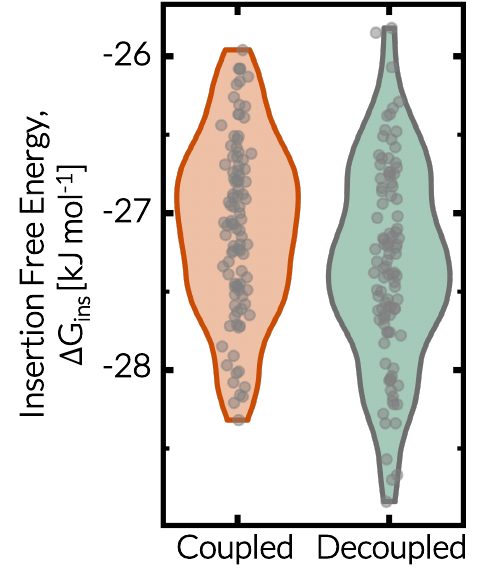}
    \caption{\label{fig.8}
    Violin-plot comparison of water insertion free energies in pure water, computed using coupled and decoupled alchemical pathways.
    }
\end{figure}

Finally, we emphasize that the discrepancy between coupled and decoupled pathways is specific to electrolyte solutions and is absent in pure water. Figure~\ref{fig.8} compares the insertion free energies obtained from the same coupled and decoupled protocols in the absence of ions. The two protocols yield statistically consistent chemical potentials, with mean values of
$-27.09 \pm 0.56~\mathrm{kJ\,mol^{-1}}$ and
$-27.36 \pm 0.59~\mathrm{kJ\,mol^{-1}}$ for the coupled and decoupled pathways, respectively. Each violin visualizes the probability distribution of $\Delta G_{\text{ins}}$ values, and the overlaid points represent the individual observations from 100 independent calculations for either pathway.

Taken together, these results indicate that the discrepancy between fully decoupled and partially or fully coupled pathways originates from unphysical water--ion contacts that arise when electrostatic interactions are introduced before a sufficiently strong van der Waals excluded volume is established. In electrolyte solutions, such pathways can stabilize unrealistically close water--chloride configurations, producing large fluctuations and biased insertion free energies. These findings underscore the need to design alchemical pathways that maintain adequate short-range repulsion whenever charged species are present, thereby avoiding artifacts associated with incomplete excluded-volume coupling.

\section{Conclusions}
\label{section:conclusions}

\noindent
In this work, we examine an often underexplored aspect of free energy calculations: how the extent and order of van der Waals and Coulombic interactions influence insertion free energies, sometimes referred to as solvation free energies, obtained from staged particle insertion. By computing the excess insertion free energy of a \textsc{Tip4p-Ew} water molecule in a 1 m aqueous KCl solution using multiple coupling protocols, we demonstrate that the concurrent activation of Lennard–Jones and electrostatic interactions toward the end of the insertion path can produce systematic errors and large statistical uncertainties. Detailed inspection of the trajectories reveals that when these interactions are turned on simultaneously, the repulsive Lennard–Jones core remains too weak to counterbalance the electrostatic attraction between chloride ions and the inserted water molecule. As a result, ion-water pairs can approach at unphysically short separations. These observations indicate that reliable free energy estimates require careful path design: in practice, it will be the safest for the Lennard–Jones interactions to become fully activated prior to introducing the Coulombic part of the potential.

Free energy calculations play a central role in modern computational materials science by linking molecular-level interactions to macroscopic measures of stability. Over the past few decades, alchemical free energy calculation methods have gained widespread popularity in both academia and industry, particularly in applications such as structure-based drug discovery.\cite{Chipot2014FrontiersSystems, Hansen2014PracticalReview} This growing interest has been driven by three major developments. First, methodological advances have improved the robustness and usability of free energy calculations.\cite{Klimovich2015AGROMACS} Second, these techniques have been incorporated into widely used molecular simulation packages and tools-- including \textsc{Lammps},\cite{ThompsonComputPhysCommun2022} \textsc{Gromacs},\cite{ABRAHAM201519} \textsc{Amber},\cite{CaseJChemInfModel2025} \textsc{Gromos},\cite{ScottJPCA1999} and Py\textsc{Mol}-- making them accessible to a broad user community.\cite{Klimovich2015AGROMACS, Lashkov2021PyFepRestr:Binding, Moore2023AutomatedG, Robo2023FastSampling} Third, steady increases in computational power have enabled the application of free energy methods to increasingly complex systems.\cite{Rick2006IncreasingReweighting, Novack2025MassivelyMiniproteins} Despite extensive documentation and best-practice guidelines, however, free energy calculations remain challenging in practice, and considerable confusion persists regarding appropriate protocols and methodological choices. Electrolyte systems present additional challenges, due to the long-range nature of electrostatic interactions, which can result in a wide variety of artifacts.\cite{Shoemaker2022IdealMembranes, Shoemaker2024IdealTransport, KhalifaJPCB2025} Our work demonstrates the need for extra care and caution in applying alchemical approaches to ionic systems.

This observation presents an interesting dilemma. By construction, alchemical techniques such as the staged insertion method generate intermediate configurations that are not physically realizable. Under normal circumstances, such configurations are gradually “healed’’ as the coupled Hamiltonian approaches the physically realistic potentials of the initial and final states. In the case of end-coupled paths considered here, however, the ion-water pairs that arise possess interaction energies so large that they cannot properly relax as the interactions are fully activated. Interestingly, the corresponding pathways remain reversible and do not exhibit noticeable hysteresis, nor does the anomaly disappear with extended sampling or with finer discretization in the $\bm\lambda$ space. We cannot rule out the disappearance of such artifacts in the limit of infinite sampling and infinitely fine discretization. From a practical standpoint, however, it is clear that end-coupled paths are highly inefficient-- even if they are not fundamentally flawed.

Our finding that these anomalies arise only in the presence of ions and disappear in pure water suggests that they are likely limited to systems containing ionic species, such as electrolytes, ionic liquids, and charged polymeric or biomolecular solutions. We further expect these artifacts to be attenuated-- if not fully eliminated-- in systems containing polyatomic ions, whose constituent atoms often carry partial charges of opposite sign, thereby moderating excessively strong electrostatic attractions. Although a systematic assessment of the persistence and extent of such artifacts in solutions and melts of polyatomic ions would be valuable, from a practical standpoint it remains advisable to avoid coupled activation paths due to the risk of introducing the unphysical artifacts observed in this work.

This work helps establish best practices for computing insertion free energies and chemical potentials in ionic systems. In addition to being crucial for studying a wide variety of systems, such as polyelectrolytes,\cite{GhasemiAIChEJ2022} nanoporous membranes,\cite{LuoJMembSci2011, KolevJPCB2015, MalmirMatter2020, ShoemakerACSNano2024} porous electrodes,\cite{MerletNatMater2012, ScalfiAnnuRevPhysChem2021, UralcanACSAppMaterInterfaces2022, DoganJPCL2025} and polymeric solutions,\cite{KnoppMacromolecules1997, SauterJCTC2016} accurate evaluation of such quantities is key for the parameterization of coarse-grained\cite{CaceresPCCP2021, ShinkleJCTC2024} and implicit-solvent\cite{RckenJCP2024} models, where consistent treatment of insertion free energies is essential for thermodynamic fidelity.

\begin{acknowledgments}
\noindent
We thank A.Z. Panagiotopoulos  for useful discussions. This work was supported by the Alfred P. Sloan Foundation.  B.U. acknowledges the Anderson Postdoctoral Fellowship from Yale University. These calculations were performed at the Yale Center for Research Computing.  This work used Stampede through allocation CHM240063 from the Advanced Cyberinfrastructure Coordination Ecosystem: Services \& Support (ACCESS) program, which is supported by NSF Grants \#2138259, \#2138286, \#2138307, \#2137603, and \#2138296.
\end{acknowledgments}

\bibliography{references_abbreviated.bib}

\end{document}